\documentstyle[aps,prb,multicol]{revtex}

\begin{document}
\input{epsf}
\draft
\preprint{}
\title{Coexistence of Weak Localization and a Metallic Phase in 
Si/SiGe Quantum Wells}
\author{ V. Senz,  T. Heinzel, T. Ihn, and K. Ensslin}
\address{Solid State Physics Laboratory, ETH Z\"{u}rich, 8093
Z\"{u}rich,  Switzerland\\}
\author{G. Dehlinger, D. Gr\"utzmacher and U. Gennser}
\address{Paul Scherrer Institute, CH-5234 Villigen PSI, Switzerland\\}
\date{\today}
\maketitle
\begin{abstract}
Magnetoresistivity measurements on p-type Si/SiGe quantum wells 
reveal 
the 
coexistence of a metallic behavior and weak localization. Deep in the 
metallic regime, pronounced weak localization reduces the metallic 
behavior 
around zero magnetic field without destroying it. In the insulating phase, a 
positive magnetoresistivity emerges close to B=0, possibly related to 
spin-orbit interactions. 

\end{abstract}
\begin{multicols} {2}
\newpage 
The recently discovered metal-insulator transition (MIT) in Si-MOSFETs 
\cite{Kravchenko95} has meanwhile been observed in a variety of 
material systems, such as p-type 
\cite{Hanein98a,Simmons98,Papadakis99,Hanein99} and n-type 
\cite{Hanein98b,Ribeiro99} GaAs heterostructures, Si/SiGe 
\cite{Coleridge97,Coleridge99} and AlAs\cite{Papadakis98} quantum wells. 
These experiments challenge the 
scaling theory of localization for non-interacting electrons in two 
dimensions (2D) in the weakly disordered ($k_{f}l\gg1$) regime. 
\cite{Abrahams79} 
Since then an increasing number of experiments 
\cite{Popovic97,Simonian97,Pudalov97,Kravchenko98,Yaish99,Popovic99,Brunthaler99} 
have investigated more details of this MIT. In spite of considerable 
theoretical research 
\cite{Finkelstein83,Dobrosavljevic97,He98,Castellani98,Phillips98,Si98,Ortuno99,Chakravarty99,Altshuler99,Sarma98} 
the origin of the metallic phase is still controversially discussed.\\
In high density 2D carrier systems, which can be treated as 
non-interacting,
the scaling theory of localization fits the experimental data well, 
yielding insulating behavior as one approaches the zero temperature 
limit.
But in all systems showing a MIT (with the possible exception of 
Refs.  
\onlinecite{Ribeiro99} and \onlinecite{Yaish99}), the ratio $r_{s}$ 
between carrier-carrier 
interaction energy and kinetic energy is of the order of 10, 
suggesting that interactions are driving the formation of the 
metallic 
phase and cannot be neglected when calculating corrections to the 
conductivity.
This path is followed in the majority of the theoretical 
models,\cite{Finkelstein83,Dobrosavljevic97,He98,Castellani98,Phillips98,Si98,Ortuno99,Chakravarty99} 
although several ideas not relying on strong interactions have been 
developed as well. \cite{Altshuler99,Sarma98}
Weak localization (WL) can only describe one part of the 
total conductivity correction and additional contributions such as 
particle-particle interactions, spin-orbit interactions or 
multi-subband transport, must be included.
Experimentally only the superposition of all contributions at B=0 can 
be detected.
In total, a complex conductivity behavior $\sigma(T,B)$ is expected.\\
Recent studies on the low-field magnetoresistance in the metallic 
phase have been done in Si-MOSFETs  \cite{Popovic97} and p-type 
Si/SiGe quantum wells.  \cite{Coleridge99}
 In this publication we investigate WL effects  as a  function of 
magnetic field and temperature in the regime where the system shows 
metallic behavior. The samples used in this study are p-type Si/SiGe 
quantum wells exhibiting the MIT as a function of hole density.
We find that 1. the shape of the low-field magnetoresistance in the 
metallic phase can be
well described by standard WL theory, 2. there is no indication for a 
novel dephasing mechanism in the metallic regime, 3. the 
magnitude and even the sign of the temperature dependence of the 
resistivity 
can depend on the applied 
magnetic field, and 4. a broad negative magnetoresistance develops in 
the insulating phase, with a small positive magnetoresistance 
superimposed around zero magnetic field. These
observations indicate that the resistivity in the metallic phase is 
determined by different, similarly important contributions.\\
The samples investigated in this study were grown by molecular beam 
epitaxy,
and consist of a 200{\AA} $\rm{Si_{0.85}Ge_{0.15}}$ layer surrounded 
by
undoped Si layers, a 150{\AA} B-doped Si layer with a setback of 
180{\AA}
from the well, and a 200{\AA} undoped Si cap.  The SiGe layer forms a
triangular potential well for the two-dimensional hole gas.   Due to 
the lattice
mismatch between Si and SiGe as well as due to size quantization, the 
heavy 
hole
($m_J$ = $\pm$3/2) potential is split from the light hole ($m_J$ =
$\pm$1/2) potential, and ensures that the lowest occupied bound state 
has 
heavy hole
character. The transport effective mass of this state is $m^*
\approx 0.25 m_0$, as extracted from the temperature dependence of
Shubnikov- de Haas oscillations.  Conventional Hall bar structures 
were fabricated with a source-drain length of 0.6mm and a width of 
0.2mm. The distance between the voltage probes was 0.3mm. The hole 
density 
$p$
could be tuned between $1.1\cdot 10^{11}\; cm^{-2} \le  p  \le 
2.6\cdot
10^{11}\; cm^{-2}$ using a Ti/Al Schottky gate. Transport 
measurements using standard four terminal lock-in techniques were 
performed in a pumped liquid He
cryostat, as well as in the mixing chamber of a
$^3He/^4He$ dilution refrigerator.  The mobility in these structures 
was found to increase
strongly with carrier concentration, from 1000 $cm^2/Vs$ (for $p  =
1.1\cdot 10^{11}\; cm^{-2}$) to 7800 $cm^2/Vs$ ($p = 2.6\cdot 
10^{11}\;cm^{-2}$).
Figure 1 shows a series of magnetoresistance measurements for several 
carrier densities and temperatures.
From top to bottom, the 
   carrier density decreases and the sample undergoes a transition 
from metallic to insulating behavior at B=0 as well as for small 
magnetic fields.  For large hole densities (Figs. 1 a, b), the 
   resistivity at B=0 clearly decreases with decreasing temperature, 
   indicating metallic behavior.
 Similar results have been obtained in the metallic 
   regime in Si MOSFETs,\cite{Brunthaler99} and in SiGe 
   quantum wells with fixed carrier density,\cite{Coleridge99} 
   where the authors also discuss the broad background in terms of 
   interactions. In the present paper, we focus on the evolution of 
   $\rho(B)$ as a function of $p$.
   \begin{figure} 
    \centerline{\epsfxsize=7.5cm \epsfbox{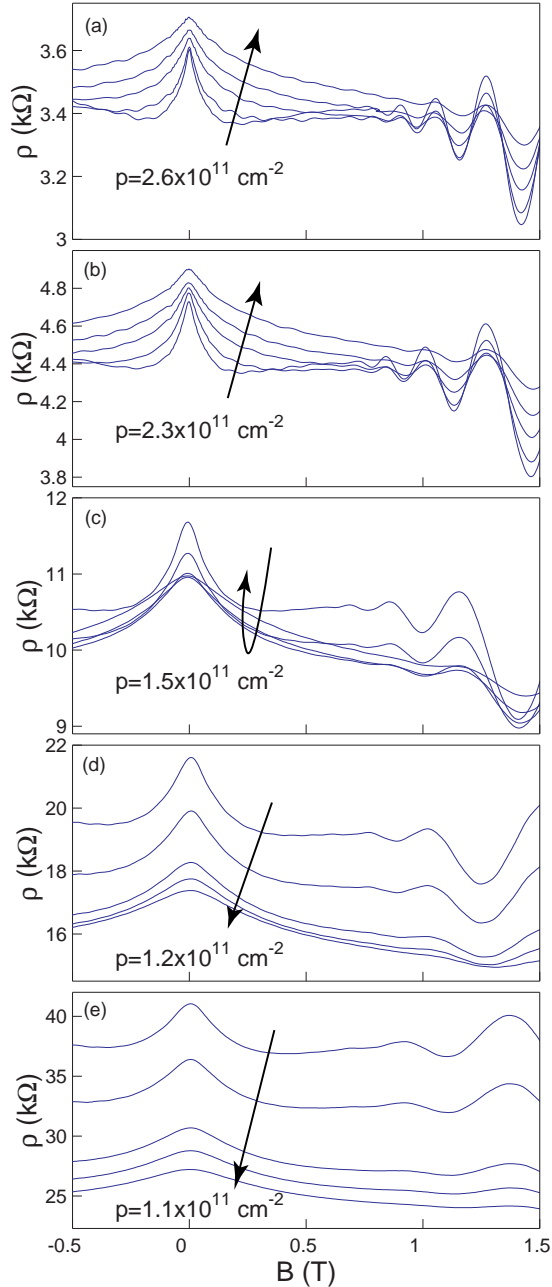}}
\caption{A series of magnetoresistivity measurements $\rho(B)$
for several carrier densities and temperatures. From top to bottom, 
the 
   carrier density decreases and the sample goes from metallic to 
   insulating behavior around zero magnetic field. The temperatures 
   for the individual traces are 0.19K, 0.36K, 0.57K, 0.74K, and 
   0.95K. The arrows denote the direction of 
   increasing temperature. In the metallic phase, the WL 
    reduces the increase of the resistivity with 
   increasing T (a, b). Close to the transition point, the behavior 
   changes from insulating to metallic as T is increased (c, see also 
   the inset in Fig. 3), while the 
   peak around $B=0$ enhances  the temperature dependence  of 
   $\rho$ in the insulating phase (d, e).}
\end{figure}
As the hole density is decreased by suitable gate 
   voltages, the metallic behavior becomes weaker (Fig.1b). 
   The sample behaves insulating as the carrier density is further 
reduced (Figs.1 d, e).  At intermediate hole densities (Fig.1c),  
$d\rho/dT<0$ at low temperatures,  but $d\rho/dT>0$ at higher 
temperatures. \\ 
   Magnetoresistivity 
   measurements  allow to distinguish different 
   contributions to the total resistivity. 
   While the WL effect leads to negative 
magnetoresistivity 
   $\rho(B)$, spin-orbit coupling results 
   in a positive magnetoresistivity.\cite{Hikami80}
   Interactions produce a complex
magnetoresisitivity, which depends on the sample 
parameters.\cite{Simonian98,Altshuler85}

From the magnetic field dependence of the 
   resistivity one can clearly discern a negative magnetoresistance 
   in the metallic phase (Figs.1 a, b). 
   Fig. 2a shows the longitudinal magnetoconductivity 
$\sigma(B)$ for $p=2.6\cdot 10^{15}m^{-2}$ around B=0 in the 
metallic phase. 
   In addition, theorectical curves for the WL correction of 
$\sigma(B)$,\cite{Hikami80} i.e. 
   \begin{equation}
\delta \sigma(B,T)=\alpha 
\frac{e^{2}}{2\pi\hbar^{2}}[\Psi(\frac{1}{2}+\frac{\tau_{B}}{2\tau_{\phi}})-\Psi(\frac{1}{2}+\frac{\tau_{B}}{2\tau_{e}})]
\label{equation1}
\end{equation}
are fitted to the data with the temperature dependent phase coherence 
time 
$\tau_{\phi}(T)$ 
   and $\alpha$ as parameters. Here, $\tau_{B}$ denotes the 
   magnetic time, $\tau_{e}$ the elastic scattering time, and $\Psi$ 
   is the digamma function. 
   The constant $\alpha$ is a 
   phenomenological parameter that describes additional mechanisms, 
for example scattering by the Maki-Thompson process,\cite{Lin84} or 
anisotropic scattering.\cite{Altshuler85}  If no such additional 
scattering mechanism exist, $\alpha$ is expected to be 
1.\cite{Fukuyama82} In n-type Si MOSFETs, 
   intervalley scattering is supposed to determine 
$\alpha$.\cite{Brunthaler99}
   Our data are fitted best for 
   $\alpha=0.61$, similar to the results of Ref. 
\onlinecite{Coleridge99}. The mechanism that leads to this reduction 
of 
   $\alpha$  remains an open question. It can not, however, 
be explained by spin-orbit scattering between the 
    light hole and the heavy hole band, since their energy separation 
    is more than $24$ meV in our system\cite{Emeleus93} and therefore 
much larger than the Fermi energy. 
 For the temperature dependence of 
   $\tau_{\phi}$, we find $\tau_{\phi}\propto T^{-\gamma}$, with 
   $\gamma =1.09\pm 0.2$ for $\alpha =1$, and $\gamma =1.29\pm 0.2$ 
for 
   $\alpha =0.61$. For dephasing by quasi-elastic 
electron-electron collisions (i.e. Nyquist noise), $\gamma$=1 is 
expected.\cite{Altshuler82}  Similar agreement 
between experiment and theory has also been 
    found in insulating 2D 
systems.\cite{Mohanty97,Aleiner98} 
    Hence, from the temperature dependence of $\tau_{\phi}$, there is 
    no 
    indication of a novel dephasing mechanism due to the presence of 
    the metallic phase. Furthermore, neither $\alpha$ nor $\gamma$ 
depend significantly on 
    $p$ in the metallic phase. \\
   Assuming that Nyquist noise causes the dephasing, we find 
   that $\tau_{\phi}$ is smaller than expected from theory, which 
states according to Ref. \onlinecite{Aleiner98}, 
   \begin{equation}
     \frac{1}{\tau_{\phi}\cdot 
T}=\frac{k_{B}e^{2}}{2\pi\hbar^{2}}\cdot\rho\cdot ln\frac{\pi 
\hbar}{e^{2}\rho}  
\label{equation2}
\end{equation}
From our fits, we find $(\tau_{\phi}\cdot T)^{-1}=3.0\cdot 
10^{11}s^{-1}K^{-1}$ (using $\alpha$=0.61), which is a factor of 
$\approx 3.2$ below the 
value expected from theory. Similar discrepancies between experiment 
and 
theory are found for insulating 2D carrier systems.\cite{Aleiner98}\\
  \begin{figure} 
\centerline{\epsfxsize=8.0cm\epsfbox{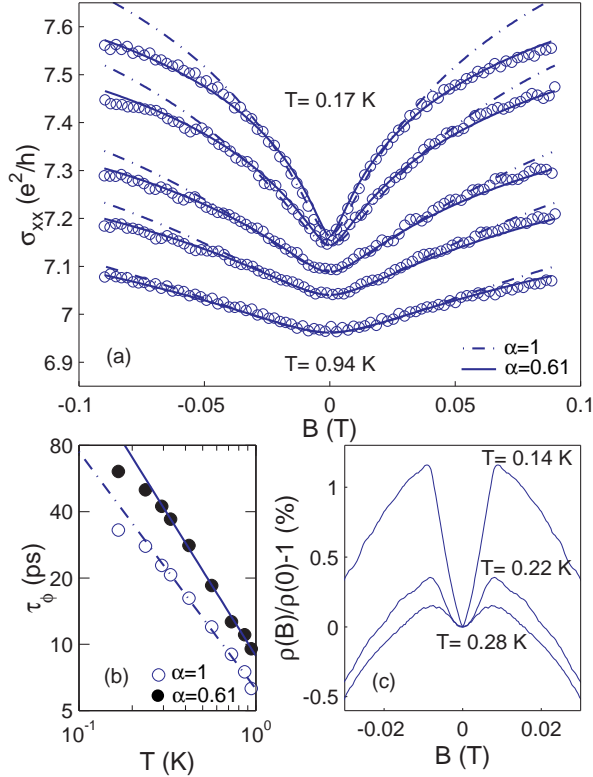}}
\caption{(a) $\sigma_{xx}(B)$ around B=0, as obtained for 
$p=2.6\cdot 10^{11}cm^{-2}$ for different temperatures (open 
circles); 
the lines represent curves according to the standard theory 
of WL, with $\tau_{\phi}(T)$ and $\alpha$ as 
parameter (see text).  
We obtain the best fits for $\alpha = 0.61$ (solid line; curves are 
fitted for -90mT$\leq$ B$\leq$ 90mT). For comparison, fits for 
-30mT$\leq$ B$\leq$ 30mT
with only $\tau_{\phi}$ as parameter (i.e. $\alpha = 1$) are shown 
(dash-dotted lines).
(b): Temperature dependence of $\tau_{\phi}$ 
as obtained from Fig. 2a,  i.e. in the metallic regime. 
We find  $\tau_{\phi}\propto T^{-\gamma}$, with 
$\gamma$ =1.09 $\pm$ 0.2 for 
$\alpha$=1 (open circles), and $\gamma$ = 1.29$\pm$ 0.2 for 
$\alpha$=0.61 (full circles). For the fits, the value for 
$\tau_{\phi}$ 
at the two lowest temperatures was left out.
(c): Relative change of $\rho(B)$ with respect to 
$\rho(0)$ for $p=1.1\cdot 10^{11}cm^{-2}$, close to $B=0$.  
For very low temperatures $T \leq 200 mK$,  
a resistance minimum occurs at $B=0$ (the data was symmetrized with 
respect to positive and negative magnetic field).}
\end{figure} These results indicate that even in the metallic regime, a 
significant 
   amount of carriers still contributes to WL. We do not find clear 
   evidence for a different dephasing mechanism than in other 2D 
   systems. Furthermore, we conclude from the existence of the WL 
peak that in our system, a spontaneous flux state at $B=0$, which 
would break the time 
reversal symmetry,\cite{Zhang94} is of minor importance.
At $B=0$ and in the metallic phase, the resistance drops 
faster 
   with decreasing temperature than the WL peak increases. 
   In order to distinguish the temperature dependence of  WL
   from the background resistance, we compare the 
   resistivity at $B=0$ with the one at  $B=0.3 T$. This field is 
larger than the characteristic field 
$B_{\tau}=\hbar/(4eD\tau)=0.11T$, and therefore the WL is quenched 
(Fig. 3). 
   Especially at low temperatures the 
   metallic behavior becomes more pronounced as one moves out of the 
WL peak. This suggests that two different contributions to the 
conductivity (or 
two conducting systems) may exist, one 
   with a metallic temperature behavior and another one with a  
   standard, insulating WL behavior. A possible theoretical 
description  could be the two-phase 
   model  proposed recently in Ref.\onlinecite{He98}.
   As one enters the insulating regime at B=0 (Fig. 1 d), a very 
broad 
   negative magnetoresistivity develops that determines the overall
   temperature dependence. In this situation, (i.e. for $k_{F}l\leq 
1$, 
   where $l$ is the elastic mean free path) $\tau_{\phi}$ 
   cannot be extracted from fitting eq. 1 to the data. 
   \begin{figure} 
\centerline{\epsfxsize=8.0cm \epsfbox{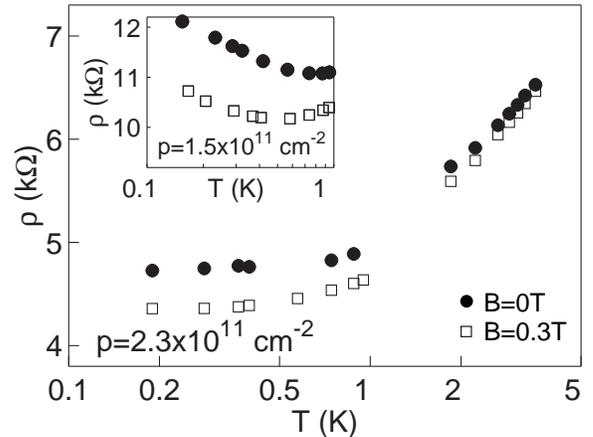}}	
\caption{$\rho$ as a function of T for the density $n=2.3\times 
10^{11}cm^{-2}$ at$B=0$ and at $B=0.3T$, where the WL 
   contribution is quenched. Inset: $\rho(T)$ at the transition 
   point from metallic to insulating behavior (Fig. 1c). }
\end{figure}  
   In this regime, the sample 
   looks rather like a conventional two-dimensional carrier gas with 
low mobility. 
 We would like to report another finding occurring in the insulating 
   phase. For very low temperatures $T \le 200$ mK and small carrier 
densities,
   an additional minimum occurs in the magnetoresistance around B=0. 
   Similar features have been observed on n-type Ga[Al]As 
heterostructures 
   \cite{Hassenkam97} and explained by spin-orbit coupling. Also, 
recent 
   data on p-type GaAs heterostructures\cite{Papadakis99} 
show a dip 
   in the magnetoresistance around $B=0$ which, however, is 
superimposed 
   on a rather flat background. Spin-orbit coupling effects are 
expected to be 
   important in p-type SiGe heterostructures 
   and could be the reason for this low-temperature feature. Note, 
   however, that in contrast to Ref. \onlinecite{Papadakis99}, we 
observe 
   this feature only deep in the {\it insulating} phase. 

 In summary, we have investigated the influence of perpendicular 
   magnetic fields on the resistance in the metallic regime of a 
two-dimensional 
   hole gas in Si/SiGe quantum wells.  A dip in the 
magnetoresistivity at B=0, possibly due to spin-orbit coupling, is 
found deep in the insulating phase. We have observed the 
   coexistence of WL and metallic behavior. Time inversion 
   symmetry seems not to be spontaneously broken at B=0 in our 
samples.  
  The temperature dependence of the dephasing 
  time $\tau_{\phi}$ suggests that Nyquist noise determines the 
dephasing even when the sample is in the metallic phase.  
We find no significant indication that $\tau_{\phi}$ 
   behaves differently than in insulating 2D systems. Our data are 
consistent with a model based on (at least) two different 
conductivity contributions for the metallic phase.

We have enjoyed fruitful discussions with P.T. Coleridge, S.V. 
Kravchenko, D. Popovic, and F. C. Zhang. Financial support from ETH 
Z\"urich and the Schweizerischer Nationalfonds is 
gratefully acknowledged.
    
\narrowtext
 
\end{multicols}
\end{document}